# The Mapping of Simulated Climate-Dependent Building Innovations

A.W.M. (Jos) van Schijndel
*Department of the Built Environment, Eindhoven University of Technology, Netherlands*

ABSTRACT: Performances of building energy innovations are most of the time dependent on the external climate conditions. This means a high performance of a specific innovation in a certain part of Europe, does not imply the same performances in other regions. The mapping of simulated building performances at the EU scale could prevent the waste of potential good ideas by identifying the best region for a specific innovation. This paper presents a methodology for obtaining maps of performances of building innovations that are virtually spread over whole Europe. It is concluded that these maps are useful for finding regions at the EU where innovations have the highest expected performances.

## 1 INTRODUCTION

Due to energy efficiency, there exist a lot of studies on innovative buildings systems. The performances of these innovations are mostly very dependent on the external climate conditions. This also means that a high performance of a specific innovation in a certain part of Europe, does not imply the same performances in other regions. Similar, innovations that did not perform very well due to local climate conditions, and therefore not commercialised, could still perform quite well in other climates. The latter can be seen as 'wasted' innovations. The mapping of simulated building systems performances at the EU scale could prevent this wasting of potential good ideas by identifying the best region for a specific innovation. This paper presents a methodology for obtaining maps of performances of building systems innovations that are virtually spread over whole Europe. Due to the novelty of the methodology it was quite difficult to find relevant references including both building simulation as well as EU mapping. Therefore the literature on both topics is presented separately. See the next Section and Section 2.2.

### 1.1 *Related work on maps*

In this Section we will focus on two important building related research areas where EU mappings are already common techniques. First, we start with cultural heritage and climate change. Grossi et al. (2007) are using maps to visualize the prediction of the evolution in frost patterns due to climate change during the 21$^{st}$ century and the potential damage to historic structures and archeological remains in Europe. Figure 1 shows an exemplarily result of the application of a freezing event map.

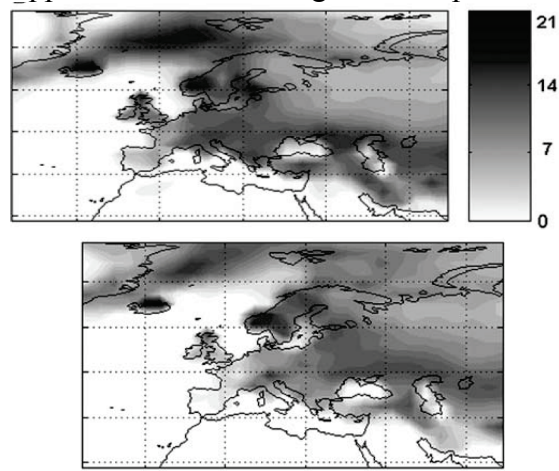

Figure 1. Pan-European maps of average yearly freezing events in 30 years period 1961–1990 (top) and far future 2070–2099 (bottom) by Grossi et al. (2007).

Similar maps as presented in Figure 1 are used to show the expected reduction of freezing and lowering the potential for frost shattering of porous building stone. The underlying data for these maps are based on regional climate models. This is the second research area where EU maps are commonly used. There is an enormous amount of literature on climate change and mapping. Therefore we illustrate the use of these maps by one state of the art regional climate model: REMO (Jacob 1997, Larsen 2010). Figure 2 shows the twenty-five-year mean modeled wind at 10 m height over Central Europe using REMO with a 10 km resolution.

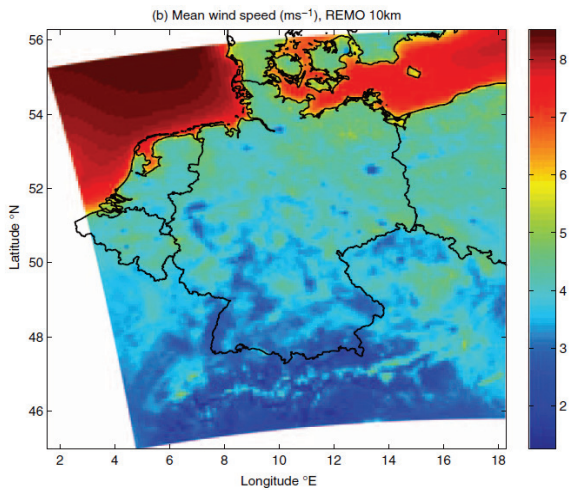

Figure 2. Twenty-five-year mean modeled wind at 10 m height over the entire domain REMO 10 km resolution (Larsen 2010)

Maps like figure 2 are suited for wind energy assessment application in Northern Europe. Moreover, literature of the related work shows that a lot of EU maps of external climate parameters are available.

## 1.2 Goal and Outline

The maps presented in the previous Section are all based on external climate parameters. However, the goal of this work is to produce maps of indoor climate related building performances. The outline of the paper is as follows: Section 2 presents the methodology for obtaining maps of performances of similar buildings that are virtual spread over whole Europe. It provides a benchmark of the EU mapping of the Bestest building. The produced maps are useful for analyzing regional climate influence on building performance indicators such as energy use and indoor climate. Section 3 presents a methodology to produce maps of systems innovations using state-space models based on a commercial case study. In Section 4, the conclusions and future research are provided.

## 2 CREATING MAPS OF BUILDING INNOVATIONS USING HAMBASE

The methodology used for obtaining the required simulation results and maps can be divided into three steps. These are presented in the following Sections.

### 2.1 External climate files

Over 130 external hourly-based climate files were produced using commercially available software (Meteonorm 2011) using the so-called wac format. Figure 3 presents the distribution of the locations over Europe.

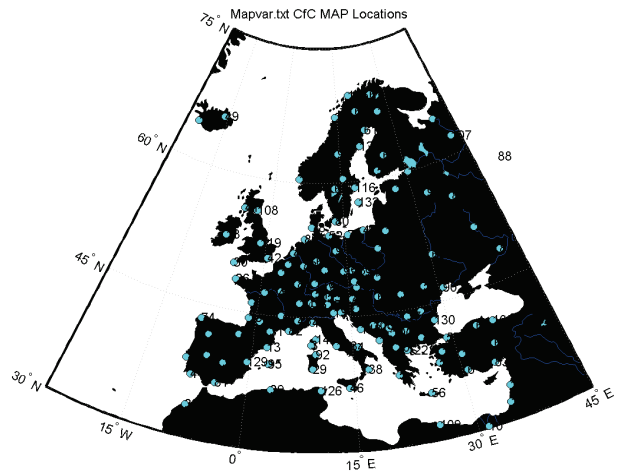

Figure 3. The distributions of the locations of the external climates in Europe.

Each climate file includes hourly based values for the common used external climate parameters: Horizontal global solar radiation [$W/m^2$] (ISGH), Diffuse solar radiation [$W/m^2$] (ISD), Cloud cover [0-1] (CI), Air temperature [$^oC$] (TA), Relative humidity [%] (HREL), Wind speed [m/s] (WS), Wind direction [0-360$^o$] (WD), Rain intensity [mm/h] (RN), Long wave radiation [$W/m^2$] (ILAH).

### 2.2 Whole building simulation model

The whole building model originates from the thermal indoor climate model ELAN which was already published in 1987 (de Wit et al. 1988). Separately a model for simulating the indoor air humidity (AHUM) was developed. In 1992 the two models were combined and programmed in the MatLab environment. Since that time, the model has constantly been improved using newest techniques provided by recent MatLab versions. The current hourly-based model HAMBase, is part of the Heat, Air and Moisture Laboratory (HAMLab 2013), and is capable of simulating the indoor temperature, the indoor air humidity and energy use for heating and cooling of a multi-zone building. The physics of this model is extensively described by de Wit (2006). The main modeling considerations are summarized below. The HAMBase model uses an integrated sphere approach. It reduces the radiant temperatures to only one node. This has the advantage that complicated geometries can easily be modeled. Figure 4 shows the thermal network. Where $T_a$ is the air temperature and $T_x$ is a combination of air and radiant temperature. $T_x$ is needed to calculate transmission heat losses with a combined surface coefficient. $h_r$ and $h_{cv}$ are the surface weighted mean surface heat transfer coefficients for convection and radiation. $\Phi_r$ and $\Phi_{cv}$ are respectively the radiant and convective part of the total heat input consisting of heating or cooling, casual gains and solar gains.

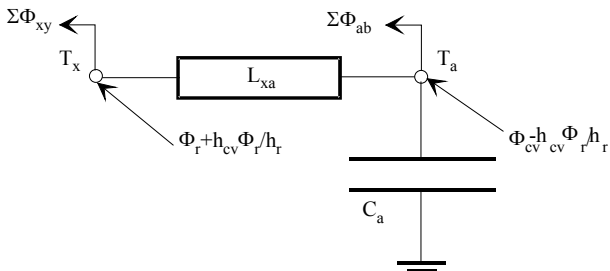

Figure 4. The room radiative model as a thermal network

The HAMBase model is adapted in such a way that all climate (.wac) files in a directory are automatically processed. For each climate file a corresponding output file is produced containing hourly based values for the indoor climate and heating and cooling power. A separate Matlab mfile is developed for calculating annual means and peak values for each location (i.e. wac file) and together with the longitude and latitude stored in a single file suitable for mapping purposes.

## 2.3 *Mapping of the results*

A MatLab mfile was developed for the visualization of the just mentioned mapping file. For the exact details of this mfile, we refer to the HAMLab website (HAMLab 2013).

## 2.4 *Mapping benchmark: Bestest using HAMBase*

The Bestest (ASHRAE, (2001)) is a structured approach to evaluate the performance of building performance simulation tools. The evaluation is performed by comparing results of the tested tool relative to results by reference tools. The procedure requires simulating a number of predefined and hierarchal ordered cases. Firstly, a set of qualification cases have to be modeled and simulated. If the tool passes all qualification cases the tool is considered to perform Bestest compliant. In case of compliance failure the procedure suggests considering diagnostic cases to isolate its cause. Diagnostic cases are directly associated with the qualification cases (Judkoff and Neymark 1995). The first qualification case, case 600 (see Figure 5) was used for the performance comparison.

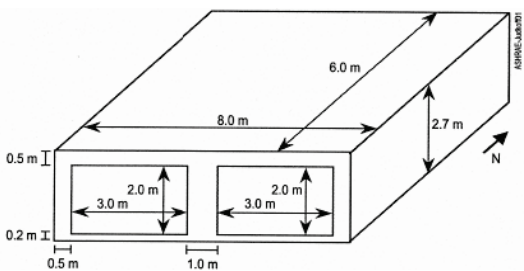

Figure 5. Bestest case 600 geometry

The thermal part of HAMBase has been subjected to a standard method of test (Bestest ASHRAE, (2001)), with satisfactory results. The accompanying climate file of the Bestest is based on weather station near Denver (USA). For further details, see Table I.

Table I Comparison of the HAMBase model with some cases of the standard test.

| Case | Nr. Simulation of | model | test min .. | max |
|---|---|---|---|---|
| 600ff | mean indoor temperature [°C] | 25.1 | 24.2 .. | 25.9 |
| 600ff | min. indoor temperature [°C] | -17.9 | -18.8 .. | -15.6 |
| 600ff | max. indoor temperature [°C] | 64.0 | 64.9 .. | 69.5 |
| 900ff | mean indoor temperature [°C] | 25.1 | 24.5.. | 25.9 |
| 900ff | min. indoor temperature [°C] | -5.1 | -6.4.. | -1.6 |
| 900ff | max. indoor temperature [°C] | 43.5 | 41.8.. | 44.8 |
| 600 | annual sensible heating [MWh] | 4.9 | 4.3.. | 5.7 |
| 600 | annual sensible cooling [MWh] | 6.5 | 6.1.. | 8.0 |
| 600 | peak heating [kW] | 4.0 | 3.4.. | 4.4 |
| 600 | peak sensible cooling [kW] | 5.9 | 6.0.. | 6.6 |
| 900 | annual sensible heating [MWh] | 1.7 | 1.2.. | 2.0 |
| 900 | annual sensible cooling [MWh] | 2.6 | 2.1 .. | 3.4 |
| 900 | peak heating [kW] | 3.5 | 2.9 .. | 3.9 |

After this verification with the Denver climate, the case 600 building model was exposed to EU climate data, located at the weather stations of Figure 3. The energy use for heating and cooling the Bestest case 600 building is shown in Figures 6 and 7. The mean annual heating and cooling power maps are in general quite as expected, i.e. high heating amounts in the north of Europe and higher cooling amounts in the South.

The use of these types of maps is published in van Schijndel and Schellen (2012). Furthermore the presented maps below can be used as a future benchmark for other modeling and mapping tools.

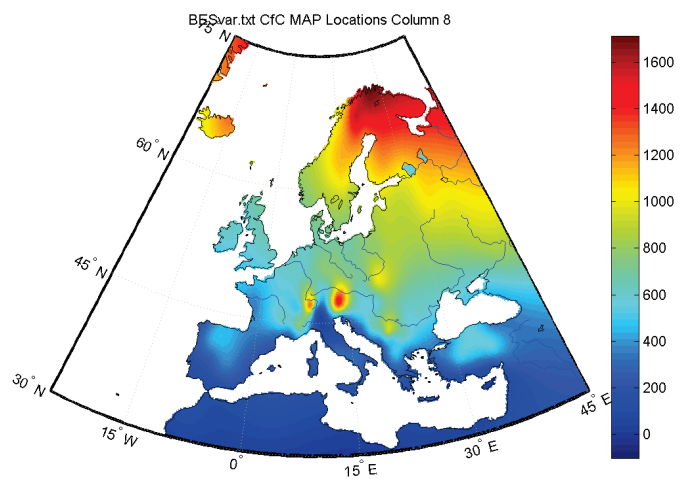

Figure 6. Mean annual heating power [W] of the Bestest case 600 building (This figure relies on color, see digital version of the paper)

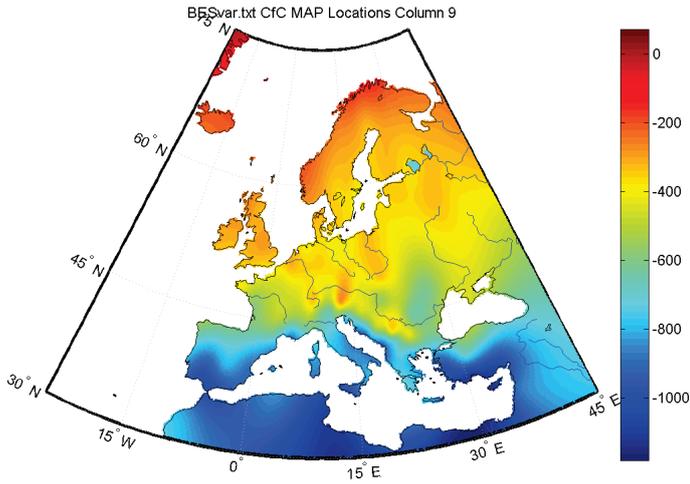

Figure 7. Mean annual cooling power [W] of the Bestest case 600 building (see also digital version of the paper for color)

So far the approach to produce maps was based on a building energy simulation (BES) tool HAMBase. In the next Section we continue with a new method (without BES) to produce maps using state space (SS) models.

## 3 CREATING MAPS OF SYSTEMS INNOVATIONS USING STATE-SPACE (SS)

A commercial case study is presented in this Section. Due to the patent protection of the industrial partner, some specific information is omitted without loss of generality. The innovation consists of a novel heat exchanger built inside a construction acting as a solar collector. Figure 8 shows the principle construction of the solar collector (in reality this is much more complicated).

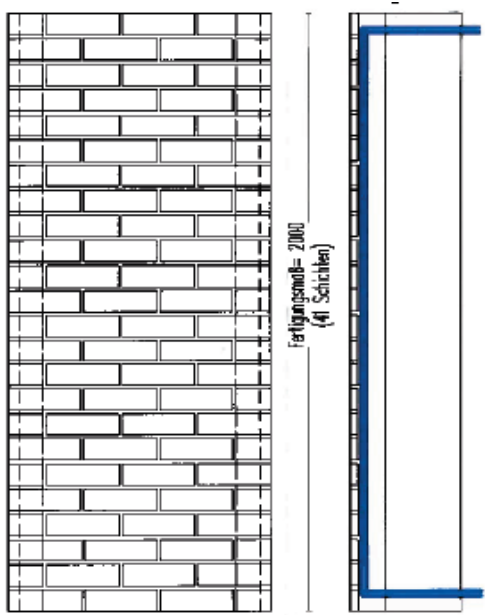

Figure 8 Construction of the solar collector

The solar collector will be used for the heating of water that directly can be used or stored for later use.

### 3.1 Modeling

A 3-State (3S) model was developed:

$$C_1 \frac{dT_1}{dt} = h\, A(\text{Tamb}(t) - T_1) - \frac{(T_1 - T_2)}{R_1} + a_I\, A\, I(t)$$

$$C_2 \frac{dT_2}{dt} = \dot{m}\, c\, (\text{Tsup}(t) - T_2) + \frac{(T_1 - T_2)}{R_1} - \frac{(T_2 - T_3)}{R_2}$$

$$C_3 \frac{dT_3}{dt} = \frac{(T_2 - T_3)}{R_2}$$

Where
*Inputs*
Tamb(t)  ambient (external) air temperature [°C]
Tsup(t)  water supply temperature [°C]
I(t)  external solar irradiance [W/m$^2$]

*States*
$T_1$  external surface temperature [°C]
$T_2$  water return temperature [°C] =
$T_3$  internal wall temperature [°C]

*Parameters:*
$\dot{m}$  water mass flow [kg/s]
c  heat capacity of water [J/kgK]
$a_I$  solar absorption factor [-] (
h  heat transfer surface coefficient [W/m$^{2\,\circ}$C]
A  surface [m$^2$]
d1  distance pipe to surface [m]
d2  distance pipe to insulation [m]
k  heat conductivity of concrete [W/mK]
$R_1$  heat resistance [ K/W] = d1/(kA)
$R_2$  heat resistance [ K/W] = d2/(kA)
$C_i$  heat capacity [J/ K]

The model was implemented using standard state-space modeling facilities of MatLab. The next Section shows the simulation and validations results.

### 3.2 Validation

Laboratory experiments were used to validate the models. All experiments were simulated using the proper parameters and boundary conditions. The results were compared in order to evaluate the predictability of the model. In Figure 8 the results for a typical experiment A is shown.

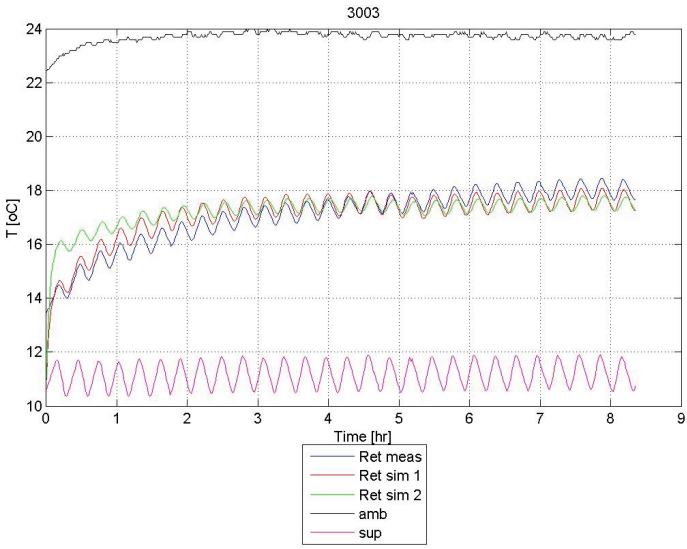

Figure 8. The simulation of experiment A: Temperatures vs time of the measured supply water(sup), the measured ambient air (amb), the simulated return water (Ret sim 1 & 2) and the measured return water (Ret).

From Figure 8 we observed that the predictability of model was satisfactory. All other tested configurations provided similar good results. Therefore we conclude that the model is quite usable for further use.

### 3.3 *Simulation using a typical Dutch climate*

The model configuration A was simulated using a reference standard Dutch climate of deBilt. Figure 9 presents the result.

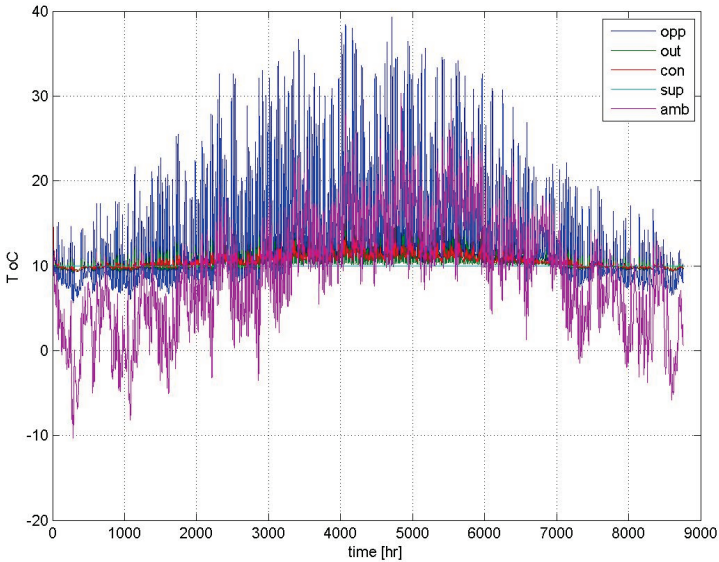

Figure 9. Simulation of model configuration A using a reference standard Dutch climate of deBilt. Temperatures versus time of the external wall surface (opp), the water return (out), the mid wall (con), the water supply (sup) and ambient air (amb).

The water supply temperature was constant held at 10 °C. The other two input signals: Ambient air temperature and solar irradiation were taken from the climate file. The main output signal is the return temperature (out). With this signal the output power can be calculated. This is shown in the next Section.

### 3.4 *Performance evaluation*

Figure 10 shows details of the model A configuration performance results.

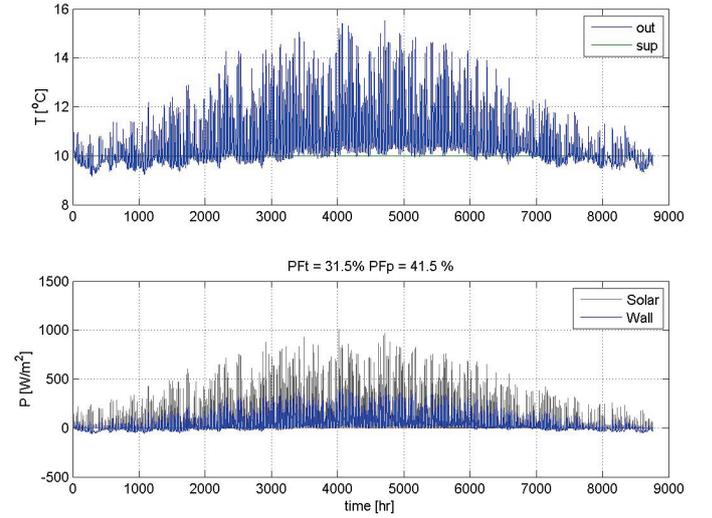

Figure 10. Performance evaluation. Top: The simulated supply and return water temperatures versus time. Bottom: The heat flux [W/m$^2$] of the incoming solar irradiation (Solar) and simulated output flux of the wall. 31.5 % of the year the wall system can be operated (PFt) The yearly mean efficiency is 41.5 % (PFp).

The output flux Pout is calculated by:

$$P_{out}(t) = \dot{m} \cdot c \cdot (T_{ret}(t) - T_{sup}(t))/A \quad [W/m^2]$$

The overall performance is evaluated as follows: Firstly, P50(t) is defined as Pout(t) with a threshold of 50 W/m$^2$. Below 50W/m$^2$, the water return temperature drops below 10.7 °C and the wall system is too inefficient. For these values P50(t) = 0. Secondly, two performance (PF) indicators are defined as follows:

PFt = Percentage of time of Pout(t) above threshold of 50 W

i.e. percentage of time of possible operation [%].

PFp = 100* sum(P50(t)) / sum(I(t)) ,

i.e. the yearly mean efficiency [%]

From Figure 10 it follows for configuration A, PFt=31.5% and PFp=41.5%.

The main parameter that affects the simulated performances is the mass flow of the water. Figure 11 provides the simulated performances PFt and PFp as functions of the mass flow.

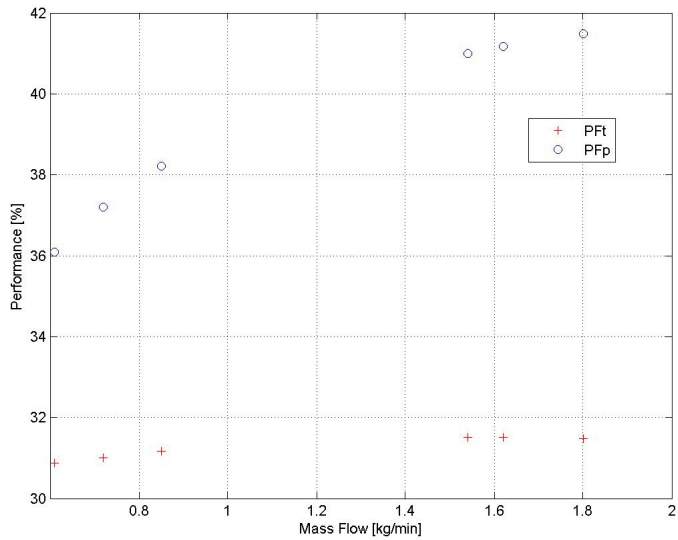

Figure 11. The simulated performances versus the mass flow.

Figure 12 presents the influence of the pump energy and surface heat transfer coefficient.

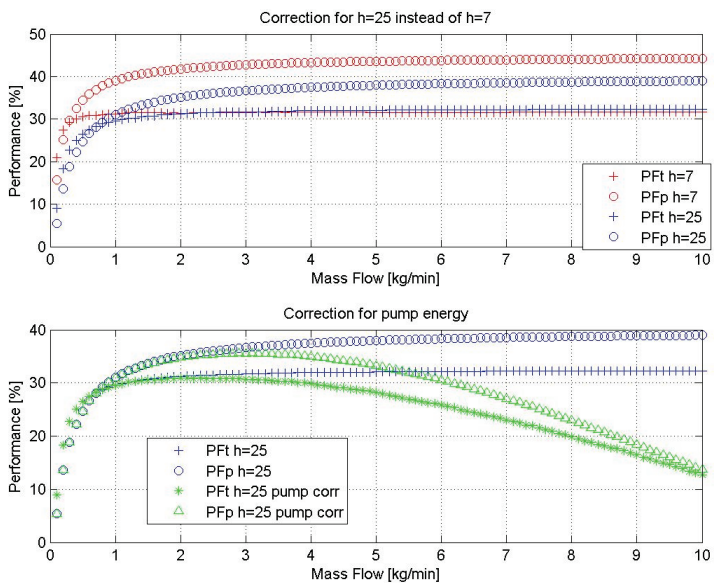

Figure 12. Influence of the pump energy and surface heat transfer coefficient. Top: The influence of a change in heat transfer surface coefficient. Bottom: Correction of the performances using pump energy.

For further simulations a more realistic surface heat transfer coefficient of 25 W/m$^2$K is used instead of 7 W/m$^2$K from the indoor experiment. The latter (i.e. h=7 W/m$^2$K) was used for the validation of the experiments. Furthermore, for the water mass flow, values between 0.2 and 2 l/min are used.

### 3.5 *Parameter study*

The following parameters were varied for the parameter study:
  * The distance from the pipe to the surface (default 35 mm) was varied: 20, 35 and 50 mm.
  * The mass flow (default 1 kg/min) was varied: 0.5, 1 and 2 kg/min.

The results are shown in Table II and III.

*Table II. Efficiency Performance*

| Simulated yearly mean efficiency PFp [%] | | | |
|---|---|---|---|
| | d=20 mm | d=35 mm | d=50 mm |
| MF=0.5 kg/min | 30.6 | 24.7 | 20.2 |
| MF= 1 kg/min | 39.0 | 30.9 | 25.2 |
| MF= 2 kg/min | 44.3 | 34.8 | 28.0 |

*TableIII3. Operation Time Performance*

| Simulated Operation time PFt [%] | | | |
|---|---|---|---|
| | d=20 mm | d=35 mm | d=50 mm |
| MF=0.5 kg/min | 29.8 | 26.5 | 23.7 |
| MF= 1 kg/min | 33.1 | 29.5 | 26.5 |
| MF= 2 kg/min | 34.5 | 30.9 | 27.7 |

The optimal efficiency performance for a Dutch climate is 44.3% with the accompanying mass flow of 2 kg/min and pipe depth of 20 mm.

### 3.6 *EU Mapping of the standard configuration*

By replacing the Dutch climate with the climates of weather stations presented in Figure 3, it is quite easy to simulate the response of the system to each external climate. From the responses the performance indicators can be calculated (See previous Section). The results of the standard wall performances are shown in Figures 13 and 14. These results are still based on the standard wall configuration A.

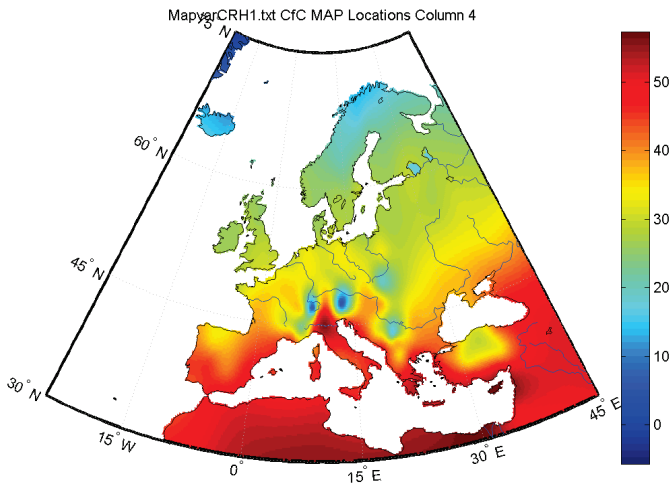

Figure 13. Efficiency (PFp) of the standard wall configuration.

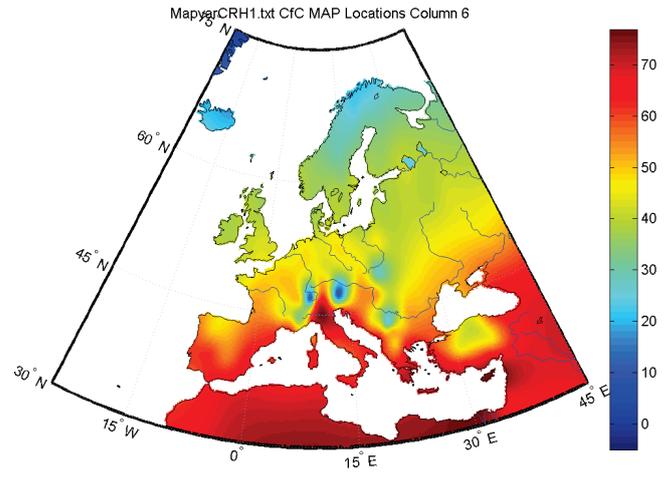

Figure 15. Optimized wall configuration Efficiency (PFp).

From figure 10 it can be seen that large parts of Europe have efficiencies of at least 45%.

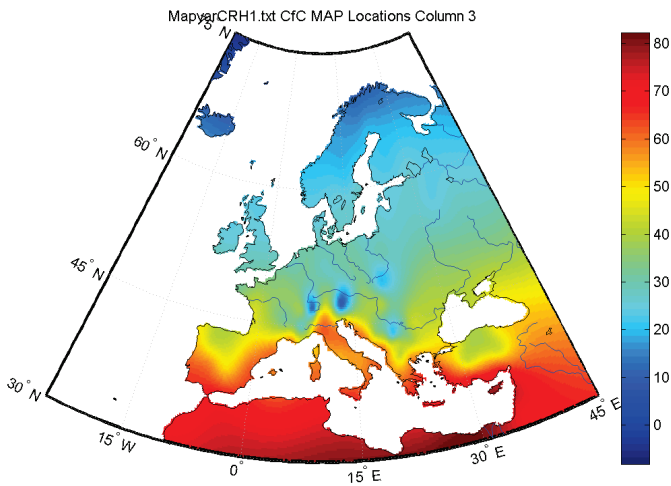

Figure 14. Percentage of time operation (PFt) of the standard wall configuration.

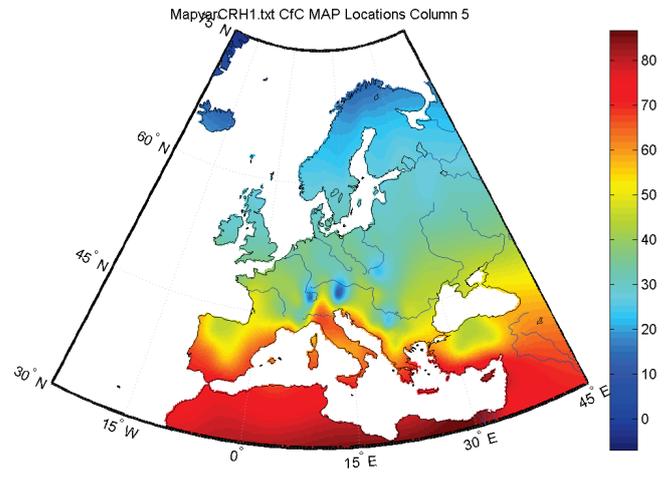

Figure 16. Optimized wall configuration Percentage of time operation (PFt).

### 3.7 *Simulation of optimized wall configurations*

All nine configurations of the parameter study (see Table II and III) were also simulated on the EU scale. For each weather station the best configuration out of nine was selected. These optimized wall configuration performances are presented in Figures 15 and 16

From figure 16 it can be seen that the areas near the Mediterranean have percentages of time of operation above 60%. The latter means that the wall collector is also operational during parts of the night.

## 4 CONCLUSIONS

*Mapping using HAMBase*
The produced maps are useful for analyzing regional climate influence on building performance indicators such as energy use and indoor climate. This is shown using the Bestest building as a reference benchmark. An important application of the mapping tool is the visualization of potential building measures over the EU. Also the performances of single building components can be simulated and mapped. It is concluded that the presented method efficient as it takes less than 15 minutes to simulate and produce the maps on a 2.6GHz/4GB computer. Moreover, the approach is applicable for any type of building

*Mapping using state space*
The main objective was to simulate and optimize the thermal performance of innovative solar collector under different EU climate conditions using state space modeling:
(1) The solar collector was successfully modeled;
(2) The validation of this model using existing measurements was satisfactory;
(3) The solar collector model was successfully simulation using 130 EU weather stations;
(4) Estimation of minimal and maximal performance was done by a parameter study;
(5) EU Maps of the performance were created.

*Regarding the EU performance of the solar collector*
Large parts of Europe have solar collector efficiencies of at least 45%, the exact details are provided in Figure 15. Furthermore, areas near the Mediterranean have percentages of time of operation above 60% (exact details are shown in Figure 16). The latter means that the solar collector is even operational during parts of the night. It is concluded that this study shows that the solar collector could be applicable in large parts of Europe. However, the reader should notice that the solar collector simulation results in this study are based on two assumptions: The supply water temperature is constant at 10 $^oC$ and all heat produced by the wall collector is usable at any time. Under most circumstances this is not very realistic. Therefore it is recommended to include buildings, systems and controllers details into the modeling for more realistic performance simulations and design of promising integrated configurations.

## 5 FUTURE RESEARCH

*Including future climates*
Within the mentioned EU FP7 project 'Climate for Culture', detailed EU external climate files are currently under development for the period 1960 – 2100 using the REMO model (Jacob et al. 1997) and a moderate climate scenario. With these future external climate files we will be able to predict future building performance indicators. Together with the EU mapping tool this could be helpful to locate EU regions with the highest impact on the specific building performances.

*Towards a state space based mapping tool*
Currently we are working on a more general state space mapping tool in MatLab. This tool will become public available.